\documentclass[usenatbib]{mn2e}
\usepackage{epsfig}

\newcommand{\etal}{et~al.} 

\newcommand{\neuthy}{H{\sc i} }
\newcommand{\kms}{$\mbox{km~s}^{-1}$}
\newcommand{\msol}{$\mbox{M}_\odot$}
\newcommand{\lsol}{$\mbox{L}_\odot$}

\newcommand{\specdfig}[2]        
{
   \begin{center}
     \begin{minipage}[t]{0.45\textwidth}
         \psfig{file=#1.eps,height=0.65\textwidth,width=1\textwidth,angle=0}
     \end{minipage}
     \hfill
     \begin{minipage}[t]{0.45\textwidth}
         \psfig{file=#2.eps,height=0.65\textwidth,width=1\textwidth,angle=0}
     \end{minipage}
   \end{center}
}

\newcommand{\specsfig}[1]        
{
   \begin{center}
     \begin{minipage}[t]{0.45\textwidth}
         \psfig{file=#1.eps,height=0.65\textwidth,width=1\textwidth,angle=0}
     \end{minipage}
   \end{center}
}

\begin{document}

\title[Masers in the LMC]{Masers associated with high-mass star formation regions in the Large Magellanic Cloud}
\author[Ellingsen \etal]{S.\ P. Ellingsen,$^{1,2}$\thanks{Email: Simon.Ellingsen@utas.edu.au} S.\ L. Breen,$^{1,3}$ J.\ L. Caswell,$^3$ L.\ J. Quinn,$^{4}$ G.\ A. Fuller$^{4}$\\
  \\
  $^1$ School of Mathematics and Physics, University of Tasmania, Private Bag 37, Hobart, Tasmania 7001, Australia\\
  $^2$ Max Planck Institut f\"ur Radioastronomie, Auf dem H\"ugel 69, 53121 Bonn, Germany\\ 
  $^3$ Australia Telescope National Facility, CSIRO, PO Box 76, Epping, NSW 1710, Australia\\
  $^4$ Jodrell Bank Centre for Astrophysics, Alan Turing Building, School of Physics and Astronomy, The University of Manchester,\\ Oxford Road, Manchester M13 9PL}

 \maketitle
  
 \begin{abstract}
  
We report the results of a sensitive search for 12.2-GHz methanol maser emission towards a sample of eight high-mass star formation regions in the Large Magellanic Clouds which have been detected in other maser transitions.  We detected one source towards the star formation region N105a. This is the first detection of a 12.2-GHz methanol maser outside our Galaxy.  We also made near-contemporaneous observations of the 6.7-GHz methanol and 22-GHz water masers towards these sources, resulting in the detection of water maser emission in six new sources, including one associated with the strongest 6.7-GHz maser in the Magellanic Clouds {\em IRAS}\,05011-6815.  The majority of the maser sources are closely associated with objects identified as likely Young Stellar Objects (YSO) on the basis of {\em Spitzer Space Telescope} observations.  We find that the YSOs associated with masers tend to be more luminous and have redder infrared colours than the sample as a whole.  SED modeling of the YSOs shows that the masers are associated with sources of higher central mass, total luminosity and ambient density than the majority of YSOs in the LMC.  This is consistent with the well-established relationship between luminous methanol and water masers and young, high-mass objects observed in the Galaxy. 

\end{abstract}


\section{Introduction}

The Magellanic Clouds represent a unique laboratory for studying star formation in an environment which differs from that of our Galaxy.  They are nearby (in comparison to most other galaxies), at distances of approximately 50 and 60~kpc respectively for the Large and Small Magellanic Clouds \citep{Feast99,Walker99}.  This enables them to be studied with intrinsically high sensitivity and resolution.  The relative star formation rates in both the Large Magellanic Cloud (LMC) and the Small Magellanic Cloud (SMC) are greater than in the Milky Way, but the metallicity is significantly lower \citep{Russell+92}.  These conditions are thought to be similar to those in larger galaxies earlier in the history of the Universe.  Hence studies of star formation in the Magellanic Clouds can yield insights into star formation at redshifts of $z > 1$.

Interstellar masers are excellent signposts of high-mass star formation in the Milky Way.  They are intense and occur at wavelengths that are not significantly absorbed by intervening clouds of gas and dust \citep{Green+09}.  The first complete survey of the LMC and SMC was recently undertaken in the 6.7-GHz transition of methanol and the 6.035-GHz transition of OH \citep{Green+08}.  The survey, made with the methanol multibeam receiver on the Parkes telescope, had a 1-$\sigma$ sensitivity of 0.13~Jy and 0.09~Jy over the entire SMC and LMC respectively (the extent of both Clouds, as defined by the CO and/or \neuthy\/ emission).  The region of the LMC containing the bulk of the CO emission and all the previously known masers were surveyed to a better 1-$\sigma$ sensitivity of approximately 60~mJy.  Despite this comprehensive and sensitive search only four 6.7-GHz methanol masers have been discovered in the LMC \citep{Sinclair+92,Ellingsen+94,Beasley+96,Green+08}.  Taking into account the relative rates of high-mass star formation in the two systems, this is a factor of 4--5 lower than expected on the basis of the observed luminosity distribution of Galactic 6.7-GHz methanol masers \citep{Green+08}.  In contrast, the number of detected ground-state OH and water masers is roughly consistent with the difference in the star formation rates between the Milky Way and LMC.  This discrepancy has been attributed to the lower metallicity of the LMC.

The current rate of high-mass star formation within the Magellanic clouds, and its relationship to the dynamical interaction of the clouds with the Milky Way, are a topic of active investigation.  The {\em Spitzer Space Telescope} Legacy program SAGE (Surveying the Agents of a Galaxy's Evolution) has made observations of a number of nearby galaxies (including the LMC) to examine the current star formation rate \citep{Meixner+06}.  The unique association of 6.7- and 12.2-GHz methanol masers with young, high-mass star forming regions has made them very useful signposts of such regions within our Galaxy.   A recent comparative study of 6.7- and 12.2-GHz methanol masers in the Milky Way has shown that 12.2-GHz masers are preferentially associated with more luminous 6.7-GHz sources \citep{Breen+09}.  \citeauthor{Breen+09} also used their observations and comparisons with published 1.2-mm dust continuum and mid-infrared data to start to quantify the maser evolutionary scheme proposed by \citet{Ellingsen+07}.  They found that the 12.2-GHz methanol masers are associated with a later evolutionary phase than those sources which show emission from only the 6.7-GHz transition.   The median ratio of the peak flux density of the 12.2- to the 6.7-GHz transitions was found to be 1:5.9, similar to previous studies.  It was also shown that the ratio depends on luminosity (and hence age), increasing as the luminosity of the 6.7-GHz emission increases (see Fig.~4 of \citeauthor{Breen+09})

In order to investigate whether the properties of 12.2-GHz methanol masers in the LMC differ from their Galactic counterparts we have undertaken new, sensitive observations towards known 6.7-GHz methanol and OH maser sources in the LMC.  \citet{Ellingsen+94} searched for 12.2-GHz methanol masers towards the two sources known at the time N11 (MC18) and N105a (MC23), with 3-$\sigma$ detection upper limits of 0.1 and 0.3~Jy respectively.  There are no published observations at 12.2-GHz towards the two more recently discovered 6.7-GHz methanol masers in the LMC {\em IRAS}\,05011-6815 and N160a (MC76), nor at the locations of any of the known OH or water masers.  

In contrast to methanol masers, the abundance of 22-GHz water masers in the Magellanic clouds appears to be consistent with that observed in the Milky Way.  This suggests that water masers may be more useful signposts of young high-mass star formation regions within the LMC.  In order to better characterize the properties of methanol and water masers in the LMC we have also made observations of the 6.7-GHz methanol and 22-GHz water masers in the LMC which were nearly contemporaneous with our 12.2-GHz search. 

\section{Observations \& Data Reduction} \label{sec:obs}

We have searched a sample of eight known maser regions in the LMC for the presence of 12.2-GHz methanol masers.  The sample included all known ground and excited-state OH masers, 6.7-GHz methanol masers and 22-GHz masers in the Magellanic Clouds.  Observations were carried out with the Australia Telescope National Facility (ATNF) Parkes 64-m radio telescope during 2008 June 20--25.  The observations were made with the Ku-band receiver which detected two orthogonal linear polarizations.  It had typical system temperatures of 205 and 225 Jy for the respective polarizations throughout the observations.  The Parkes multibeam correlator was configured to record 8192 spectral channels across a 16-MHz bandwidth for each polarization.  This corresponds to a velocity coverage of $\sim$290~\kms\/ and after Hanning smoothing, a spectral resolution of 0.08~\kms.  The Parkes radio telescope has RMS pointing errors of $\sim$10 arcsec and at 12.2-GHz the telescope has a half-power beam-width of 1.9~arcmin.

All sources were observed at a fixed frequency of 12167-MHz (i.e. with no Doppler tracking), which alleviated the requirement for a unique reference observation to be made for each of the source positions.  The data were processed using the same method as described in \citet{Breen+09}, except that a running median of 50 channels (as opposed to 100 channels) was used to correct the spectrum baseline.   The data were reduced using the ASAP (ATNF Spectral Analysis Package).  Alignment of velocity channels was carried out during processing.  Absolute flux density calibration was achieved by observing PKS\,B1934-638 each day which has an assumed flux density of 1.825 Jy at 12167-MHz \citep{Sault03}.  Each of the fields was observed for 160--170 minutes, resulting in a typical RMS in the final spectrum (after Hanning smoothing and averaging over polarizations and time) of 27~mJy and a corresponding 5-$\sigma$ limit of 0.135~Jy.  The adopted rest frequency was 12.178597~GHz \citep{Muller04}.

The 6.7-GHz observations reported by \citet{Green+08} were made in 2006 June and September, approximately 2 years prior to the 12.2-GHz observations reported here.  As masers are known to be variable we undertook additional 6.7-GHz observations using the Parkes 64-m radio telescope in 2008 August ($<$ 2 months after the 12.2-GHz observations) using the methanol multibeam receiver.  The observations were made in MX mode, with 10 minutes per-beam on source toward each of the four known 6.7-GHz methanol maser sites (see \citet{Green+09} for a detailed description of the system parameters).  We recorded a 4 MHz bandpass for each beam, for each of two circular polarizations for both the 6.7-GHz methanol and the 6.035 MHz OH transitions.  The spectra were produced by an autocorrelation spectrometer with 2048 spectral channels.  The data were not Hanning smoothed and the resulting velocity resolution for the 6.7-GHz observations was 0.11~\kms.  This is very similar to the velocity resolution of the 12.2-GHz observations.  A fault with one of the beams (beam 3) meant that its data had to be excluded, resulting in an on-source integration time of 60 minutes per source.

The 22-GHz water maser observations were made with the Australia Telescope Compact Array (ATCA) during 2008 August 18,19, with the array in the 6B configuration.  The observations were made in Director's time allocations and spanned approximately 7 hours on the first day and 5 hours on the second day.  A total of eight different pointing centres were observed.  These covered the locations of all the 11 previously known star-formation maser sites in the LMC.  The observing band was centred at a fixed frequency of 22213 MHz.  The correlator was configured with 1024 channels across a 16 MHz bandwidth recording a single linear polarization and also allowing measurement of continuum emission. The data were not smoothed, giving a velocity resolution of approximately 0.25~\kms\/ and a usable velocity range of approximately 195 \kms.  PKS\,B1934-638 was used as the primary flux calibrator, which at this frequency is assumed to have a flux density of 0.83 Jy \citep{Sault03}.  Observations of PKS\,B0537-441 were used to calibrate the bandpass and PKS\,B0637-752 was used as the phase calibrator.  The observations were made as a series of 90 second integrations.  These were regularly interleaved with observations of the phase calibrator.  In total, each source was observed 9 or 10 times over the two observing sessions resulting in an onsource time of 13.5 or 15 minutes per source.  We also used the line-free spectral channels from the water maser observations to search for continuum emission associated with the star formation regions.  The observations had only a 16-MHz bandwidth and thus were not very sensitive; however they do have higher spatial resolution than previous studies of these regions. The typical RMS for the continuum images was 0.2--0.3~mJy beam$^{-1}$.

\section{Results} \label{sec:results}

\subsection{12.2-GHz methanol masers}

\begin{figure}
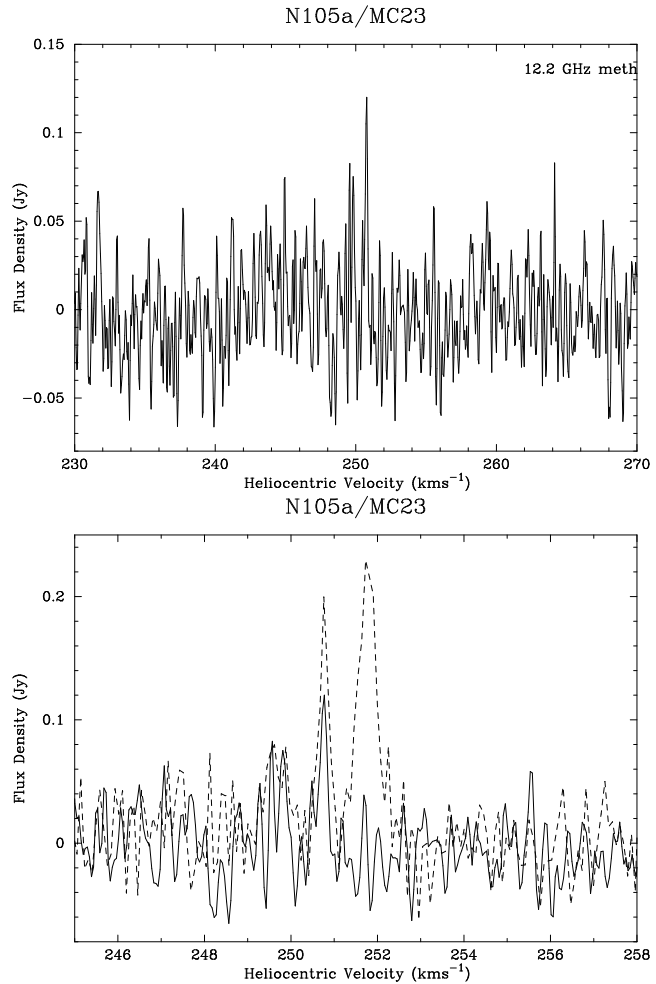

  \psfig{figure=fig1a.eps,width=8.5cm,angle=0}
  \psfig{figure=fig1b.eps,width=8.5cm,angle=0}
  \caption{The top plot shows the 12.2-GHz emission in N105a.  The maser peak is at 250.8~\kms\/ (Heliocentric velocity frame), with a peak 4.6 times than of the RMS noise in the spectrum.  The 12.2-GHz peak coincides in velocity with a secondary peak in the 6.7-GHz spectrum and a comparison of the two spectra is shown in the bottom plot with the 12.2-GHz emission shown as a solid line and the 6.7-GHz as a dashed line (NOTE: a reduced velocity range is shown on the comparison plot to allow greater detail to be seen).}
  \label{fig:n105a}
\end{figure}

The known LMC masers are contained in eight distinct fields which were targets for a 12.2-GHz methanol maser search.  Table~\ref{tab:results} lists either the detected peak flux density and velocity, or the 5-$\sigma$ noise level in the final spectrum.  A heliocentric velocity range of 120 -- 450 \kms\/ was observed for each source.  The only source detected was N105a (MC23), which coincidentally was also the first 6.7-GHz methanol maser detected in the LMC.  The 12.2 GHz spectrum of N105a is shown in the upper panel of Fig.~\ref{fig:n105a}.  The peak of the 12.2-GHz emission is 0.12~Jy (4.6-$\sigma$). As it is within the velocity range of the 6.7-GHz emission, and is also apparent when the total 12.2-GHz dataset is split in half, or when the two polarizations are examined independently, we are confident that it is a genuine detection.

Comparing our 12.2-GHz spectrum of N105a with the 6.7-GHz emission observed in 2006 June \citep{Green+08} shows that the 12.2-GHz peak coincides with the second strongest spectral peak in this source.  \citet{Breen+09} found that for approximately 75 per cent of the sources where they detected both 12.2- and 6.7-GHz emission the peak of the two spectra occurred at the same velocity.  For the remaining 25 per cent the 12.2-GHz peak coincided with a secondary 6.7-GHz feature.  Although the association of the 12.2-GHz methanol maser with a secondary 6.7-GHz spectral feature is not unusual, it is also possible that the 6.7-GHz maser may have varied in the period between the observations reported by \citeauthor{Green+08} (June 2006) and these observations.  Comparison of the August 2008 6.7-GHz methanol maser spectrum, with that shown in Fig.~3 of \citet{Green+08} shows no detectable variation over the two-year period between the observations.  The observations of the other known 6.7-GHz methanol masers in the LMC and the 6.035-GHz OH maser in N160a also show no significant changes in the absolute, or relative intensity of the maser features compared to those published by \citeauthor{Green+08} (see Figure~\ref{fig:spectra}).  

\subsection{Water maser observations} \label{sec:waterres}

\begin{table*}
\caption{A summary of the maser observations towards 12.2- and 6.7-GHz methanol masers and 22-GHz water masers undertaken as part of this project.  All velocities are given in the heliocentric frame of reference.  For the 12 sites where water masers were detected, the position is that of the water maser emission observed in our ATCA observations.  For the remaining four sites the position is that of the methanol or OH maser emission reported by \citet{Green+08}.  New detections are indicated by an asterisk after the peak flux density for that transition.  Where observations were made, but no emission was detected the 5-$\sigma$ limit for the spectra are given (ATCA spectra were formed from the image cubes).}
  \begin{tabular}{lllllllllc} \hline
      \multicolumn{1}{c}{\bf Source} & \multicolumn{1}{c}{\bf RA}  & \multicolumn{1}{c}{\bf Dec} & \multicolumn{4}{c}{\bf Methanol Masers}  & \multicolumn{3}{c}{\bf Water Masers} \\
      \multicolumn{1}{c}{\bf name}     & \multicolumn{1}{c}{\bf (J2000)} & \multicolumn{1}{c}{\bf (J2000)} & \multicolumn{2}{c}{\bf 12.2 GHz} & \multicolumn{2}{c}{\bf 6.7 GHz} & \multicolumn{2}{c}{\bf Velocity} & \multicolumn{1}{c}{\bf Peak} \\
  & & & \multicolumn{1}{c}{\bf Peak} & \multicolumn{1}{c}{\bf Velocity}  & \multicolumn{1}{c}{\bf Peak} & \multicolumn{1}{c}{\bf Velocity} & \multicolumn{1}{c}{\bf Range} & \multicolumn{1}{c}{\bf Peak} \\
  & \multicolumn{1}{c}{\bf $h$~~~$m$~~~$s$}& \multicolumn{1}{c}{\bf $^\circ$~~~$\prime$~~~$\prime\prime$} & \multicolumn{1}{c}{\bf (Jy)} & \multicolumn{1}{c}{\bf (km s$^{-1}$)} &  \multicolumn{1}{c}{\bf (Jy)} & \multicolumn{1}{c}{\bf (km s$^{-1}$)} & \multicolumn{2}{c}{\bf (km s$^{-1}$)}  & \multicolumn{1}{c}{\bf (Jy)} \\
  \hline \hline

  N11/MC18                           & 04 56 47.10 & -66 24 31.7 & $<$0.130 &              & 0.34 & 300.8 & 195--400 &             & $<$0.10      \\
  {\em IRAS}\,05011--6815 & 05 01 01.83 & -68 10 28.2 & $<$0.130 &              &  3.49 & 268.0 & 263--266 &  264.4 & 0.080$^{*}$\\
  N105a/MC23 (OH)             & 05 09 52.00 & -68 53 28.6 & $<$0.130 &             &           &            & 252--260 &  254.6 & 1.41             \\
                                                 & 05 09 52.55 & -68 53 28.0 & $<$0.130 &             &           &            & 246--268 &  266.2 & 0.76$^{*}$  \\
  N105a/MC23 (meth)          & 05 09 58.66 & -68 54 34.1 & 0.12$^{*}$ & 250.8 &  0.23 & 251.7 & 195--400 &             & $<$0.11       \\
  N113/MC24			  & 05 13 25.15 & -69 22 45.4 & $<$0.130 &             &           &            & 228--264 &  253.1 & 116.0            \\
                                                 & 05 13 17.17 & -69 22 21.6 & $<$0.260 &             &           &            & 249--286 &  257.1 & 1.96$^{*}$   \\
                                                 & 05 13 17.67 & -69 22 21.4 & $<$0.260 &             &           &            & 195--400 &             & $<$0.10       \\
   					  & 05 13 17.69 & -69 22 27.0 & $<$0.260 &             &            &            & 247--250 &  249.1 & 0.28$^{*}$   \\
  N157a/MC74			  & 05 38 46.53 & -69 04 45.3 & $<$0.135 &             &            &            & 253--271 &  269.0 & 6.5                \\
                                                 & 05 38 49.25 & -69:04:42.1 & $<$0.135 &             &            &            & 263--268 & 267.3 & 0.45$^{*}$    \\
                                                 & 05 38 52.67 & -69 04 37.8 & $<$0.135 &             &            &            & 265--269 & 266.9  & 1.4$^{*}$     \\
  N157a/MC74 (exOH)        & 05 38 45.00 & -69 05 07.4 & $<$0.135 &             &            &            & 195--400 &             & $<$0.24       \\
  N159				  & 05 39 29.23 & -69 47 18.9 & $<$0.135 &             &            &            & 243--255 &  248.2 & 1.11              \\
  N160a/MC76			  & 05 39 43.83 & -69 38 33.8 & $<$0.275 &             &            &            & 236--259 &  251.6 & 1.77              \\
 			                     & 05 39 38.97 & -69 39 10.8 & $<$0.135 &             & 0.17   & 248.3 & 252--260 &  253.1 & 2.45              \\ \hline
  \end{tabular} \label{tab:results}
  \end{table*}
  
The current water maser observations are summarized in Table~\ref{tab:results}.  The previously published searches for OH, water and methanol masers in star formation regions in the LMC span a period of more than 25 years.  As all the maser observations reported in this paper were made within a 10 week period between mid-June and the end of August 2008, they provide the first near-contemporaneous observations of the majority of star formation maser transitions in the LMC (only the ground-state OH masers were not observed).  Figure~\ref{fig:spectra} shows the emission from the 6.7-GHz methanol, 6.035-GHz OH and 22-GHz water masers detected towards the known star formation maser sites in the LMC.  This represents all the known star formation masers in the LMC with the exception of the 1.6-GHz OH transitions \citep[see][for spectra of the OH masers]{Brooks+97}.

\begin{figure*}
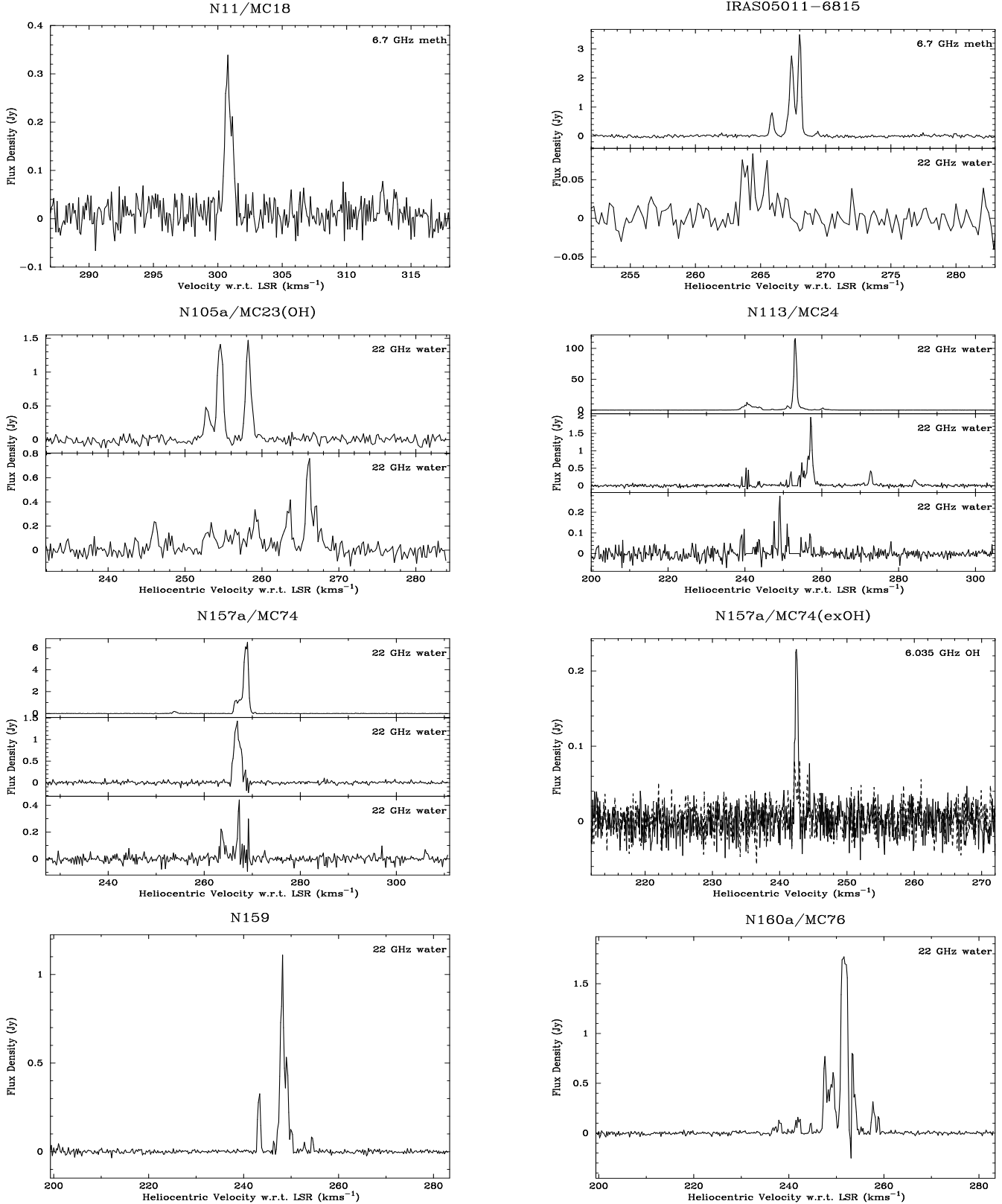

  \specdfig{fig2a}{fig2b}
  \specdfig{fig2c}{fig2d}
  \specdfig{fig2e}{fig2f}
  \specdfig{fig2g}{fig2h}
   \caption{Maser spectra of the 6.035-GHz OH, 6.7-GHz methanol and 22-GHz water masers associated with star formation regions in the LMC (except for N105a/MC23 which is shown in Figure~1).  Where multiple water masers were detected towards one region, the order of the spectra follows the order of the source positions in Table~\ref{tab:results}.  The spectra are all from the Parkes or ATCA observations described in this paper, with the exception of the 6.035-GHz OH spectrum for N157a/MC74 which is taken from \citet{Green+08}. For the 6.035-GHz OH spectra the solid line is right-hand circular polarization and the dashed line left-hand circular polarization, and the flux density scale refers to a single polarization.  NOTE: The negative spike at 253 \kms\/ towards the N160a/MC76 at $05^h39^m43.83^s$ $-69^\circ38^\prime33.8^{\prime\prime}$ is due to sidelobe interference from the nearby source at $05^h39^m38.97^s$ $-69^\circ39^\prime10.8^{\prime\prime}$.}
  \label{fig:spectra}
\end{figure*}

\begin{figure*}
  \specsfig{fig2i}
  \contcaption{}
\end{figure*}

\begin{figure*}
  \psfig{figure=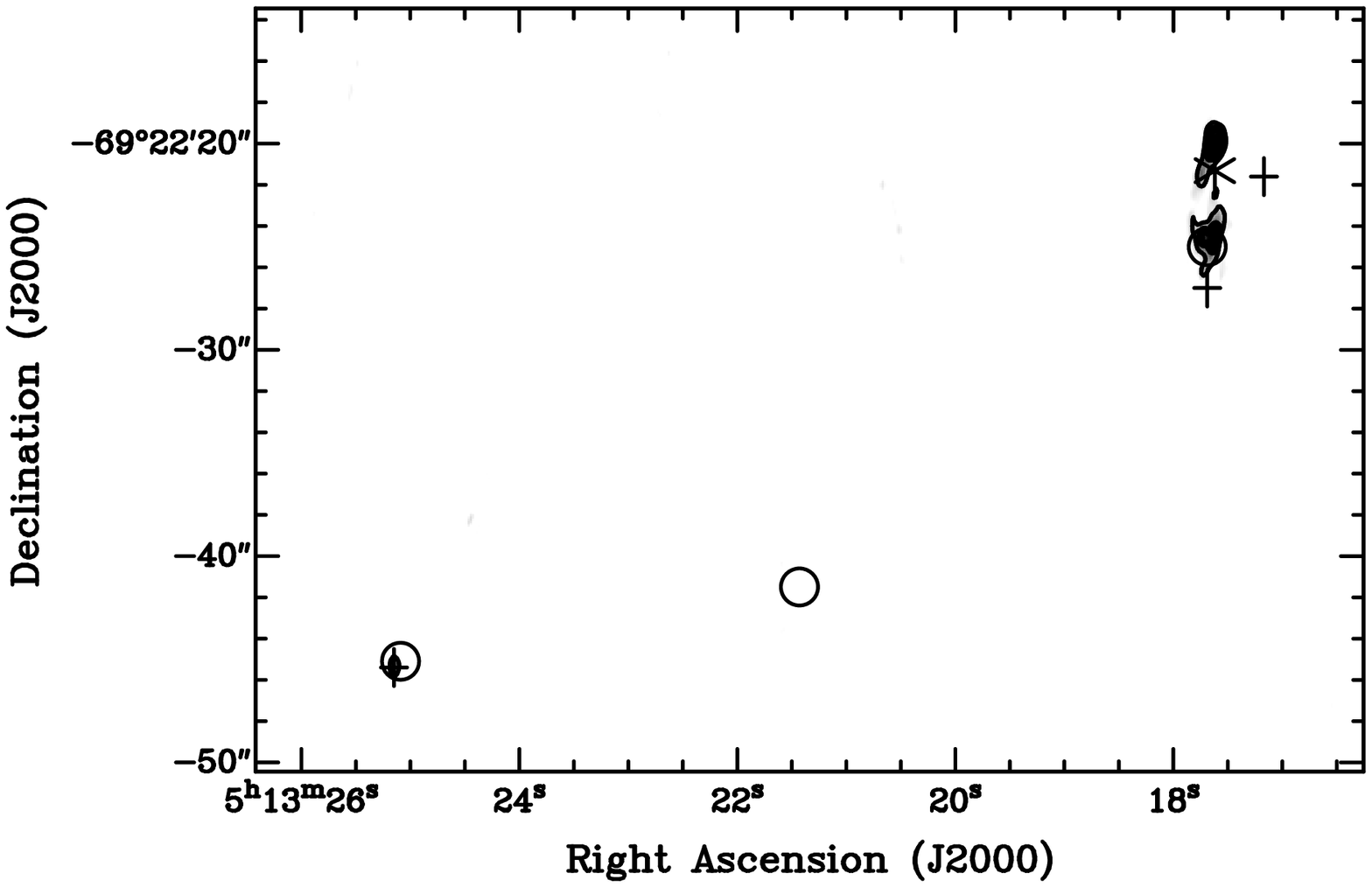,width=15cm,angle=270}
   \caption{The distribution of four water masers (crosses) in the N113 region.  The strongest water maser in the LMC is located in the bottom left of this image, while the asterisk marks the location of the maser detected by \citet{Lazendic+02} which we did not detect.  The three black circles mark the location of YSOs identified by \citet{Gruendl+09}.  The greyscale image and contours both show the 22 GHz radio continuum emission, the contours are at 3, 6 mJy beam$^{-1}$.}
  \label{fig:n113_cont}
\end{figure*}

Sensitive, high resolution water maser observations towards known or suspected high-mass star formation regions have previously been published by \citet{Lazendic+02}.  Their observations were also made with the ATCA, but using only three antennas and a maximum baseline length of 200m.  We detected all but one of the sources they observed and found six new water maser sources.  The exception was in the N113/MC24 field, where they reported the presence of a water maser at $05^h13^m17.67^s, -69^\circ22^\prime21.4^{\prime\prime}$ (J2000) with emission in the velocity range 247.3 - 254.4 \kms.  We did not detect any emission from that location, but did detect two new water masers in this field.  The first is at $05^h 13^m 17.17^s, -69^\circ22^\prime21.6^{\prime\prime}$ (J2000), offset by 2.6 arcseconds, nearly entirely in right ascension from the non-detected site.  It has a peak velocity at 257.1 \kms\/ and a much broader velocity range than any of the other sources seen in the region.  The second has a velocity range slightly lower than the non-detected source, peaking at 249.1 \kms\/ at $05^h13^m17.69^s, -69^\circ22^\prime27.0^{\prime\prime}$ (J2000) (i.e. it is six arcseconds further south, but with essentially the same right ascension).  The locations of the various masers, and infrared sources in the N113 region, along with the associated 22-GHz radio continuum emission is shown in Figure~\ref{fig:n113_cont}. We have confirmed with the authors that the position reported by \citeauthor{Lazendic+02} is correct (John Whiteoak, private communication) and it appears likely the non-detection is simply due to the variability often exhibited by water masers.  The infrared and radio continuum emission suggests that the newly detected water masers (and the non-detection) may be associated with an energetic outflow from the nearby YSO. 

The only newly detected water maser in a region where this species had not previously been detected was towards {\em IRAS}\,05011-6815 (Fig.~\ref{fig:spectra}).  This is the site of the strongest 6.7-GHz methanol maser in the LMC, which has a peak flux density of approximately 3.5~Jy.  We detect water maser emission with a peak flux density of 0.08~Jy toward the location of the methanol maser.  \citet{Lazendic+02} also searched for water maser emission towards this source, but did not detect any emission (with a limit of 0.3~Jy) at the epoch of their observations.

\citet{Oliveira+06} used the Parkes 64-m telescope to search for water masers in several regions in the LMC and identified several new water masers in the 30 Doradus (N157) region.  Our ATCA observations confirm the presence of multiple water masers in this region, although their relationship to those of \citeauthor{Oliveira+06} is not clear.  \citet{Oliveira+06} reported four masers in the N157 region, which they labelled 0539-691a, b, B and C.  We detect three masers, but all are within 35 arcseconds of the strongest maser in the region (0539-691a) previously observed by \citet{Lazendic+02}.  Their source 0539-691B lies beyond the half-power point of the primary beam of our ATCA observations, while 0539-691C is close to the edge of the beam and hence it is not surprising that we do not detect either of these sources.  The position reported by \citet{Oliveira+06} for the strongest maser 0539-691a is offset 21.3 arcseconds (18.1 arcsec later in right ascension and 11.3 arcsec further south in declination) from the position we have determined with the ATCA (which is less than an arcsecond different from the position reported by \citeauthor{Lazendic+02}).  However, applying the same pointing correction to the position reported for 0539-691b does not improve its agreement with the location of either of the new maser sources we have detected.  These differences most likely just reflect the difficulty of measuring the relative location of masers with overlapping velocity ranges with separations less than the primary beam using a single-dish telescope.



\subsection{22-GHz radio continuum emission}

\begin{figure}
  \psfig{figure=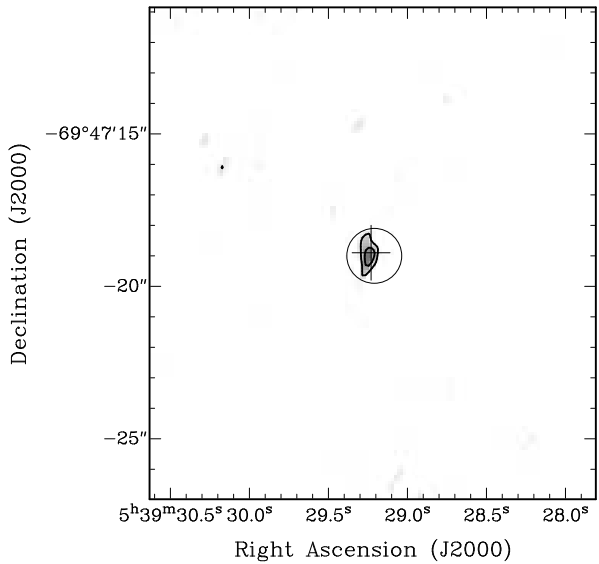,width=8.0cm,angle=270}
   \caption{The distribution of water masers (cross) and the location of the YSO (black circle) identified by \citet{Gruendl+09} in the N159 region. The greyscale image and contours both show the 22 GHz radio continuum emission, the contours are at 3 and 4 mJy beam$^{-1}$.}
  \label{fig:n159_cont}
\end{figure}

\begin{figure*}
  \psfig{figure=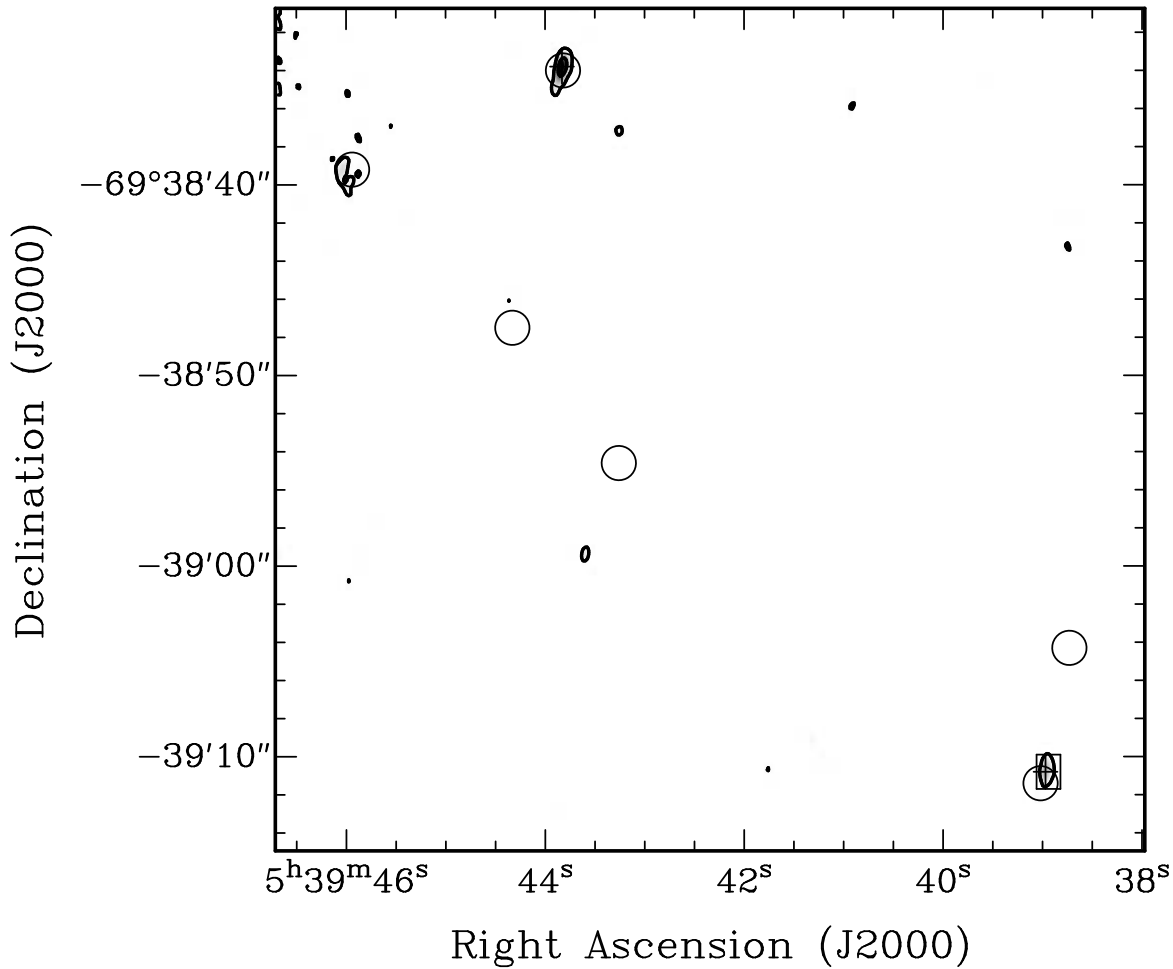,width=15cm,angle=270}
   \caption{The distribution of two water masers (crosses), and the methanol masers (square) in the N160a region. The six black circles mark the location of YSO identified by \citet{Gruendl+09}.  The greyscale image and contours both show the 22 GHz radio continuum emission, the contours are at 9 and 26 mJy beam$^{-1}$.}
  \label{fig:n160_cont}
\end{figure*}

We detected weak 22-GHz continuum emission in three fields N113/MC24, N159 and N160a/MC76, with peak fluxes of 12.9, 2.7 and 42.5 mJy beam$^{-1}$.  For N159 and N160a/MC76 the radio continuum sources are coincident with the location of the water masers to within 0.1 arcseconds.  The relative distributions of the water masers, infrared YSO and radio continuum emission are shown in Figs~\ref{fig:n113_cont} - \ref{fig:n160_cont}.  For fields where no emission was detected the 5-$\sigma$ limit was typically around 3 mJy beam$^{-1}$, with the highest being 5.2 mJy beam$^{-1}$ for N105a/MC23.  

\section{Discussion} \label{sec:discussion}

\subsection{Maser abundances in the LMC}

Previous studies have noted the apparent underabundance of methanol masers in the Magellanic Clouds in comparison to the Milky Way \citep{Beasley+96,Green+08}.  The current observations represent the first search for 12.2-GHz methanol maser towards a complete sample of 6.7-GHz methanol and 6.035 GHz OH masers.  In addition to covering all the known 6.7-GHz methanol and 6.035 GHz OH masers in the LMC, it also covers all the known ground-state OH sources, and 22-GHz water masers associated with a star formation region.

\begin{figure}
  \psfig{figure=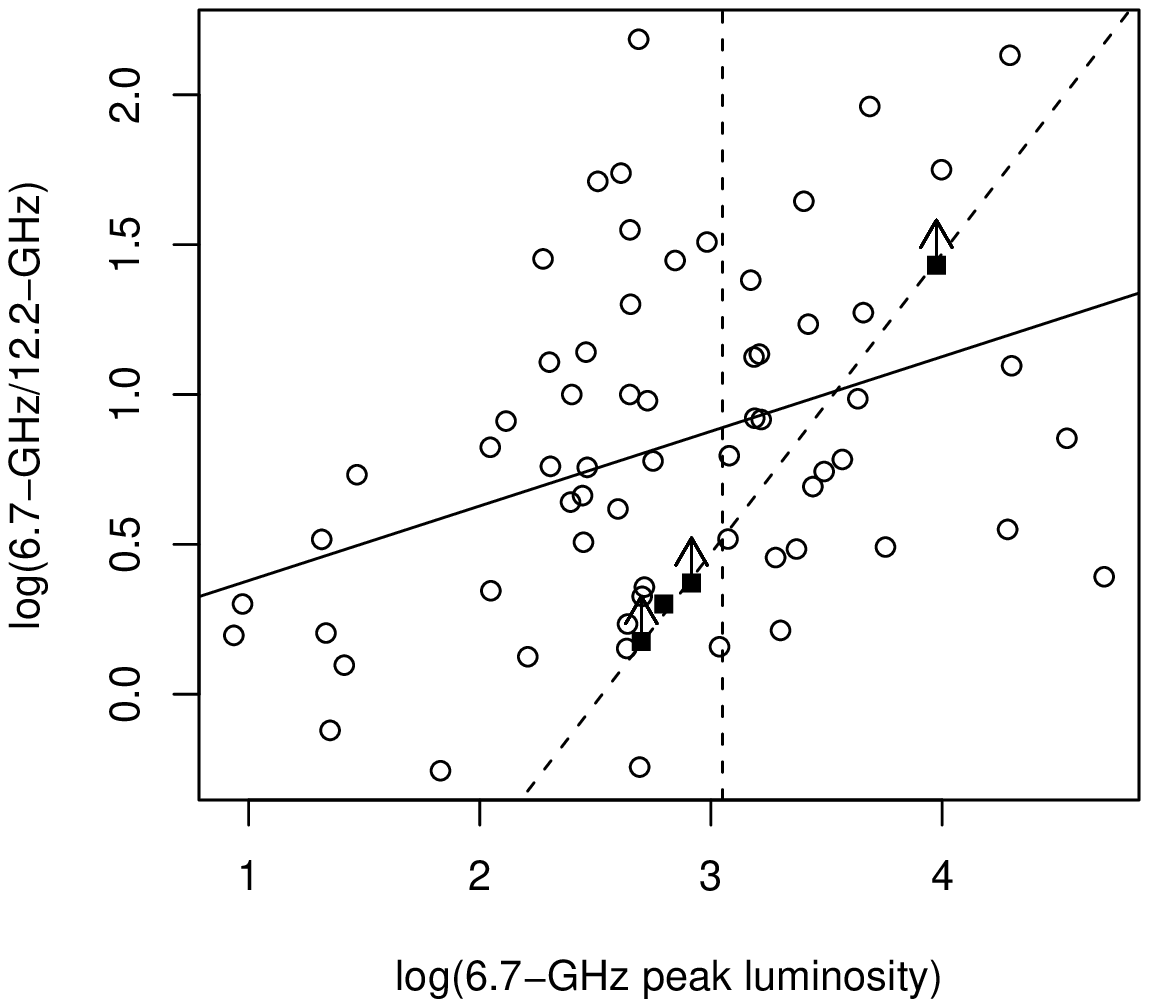,width=8.5cm,angle=270}
 \caption{A plot of the ratio of the 6.7- to 12.2-GHz peak flux density versus the peak luminosity of the 6.7-GHz methanol maser.  The open circles are Galactic sources from \citet{Breen+09}, the filled squares are the LMC measurements, with those which are lower limits (i.e. all but N105a), indicated by upward arrows.  The solid line is the linear regression relationship between 6.7-GHz peak luminosity and 6.7- to 12.2-GHz peak ratio determined by \citeauthor{Breen+09} for Galactic sources.  The dashed vertical line represents the 5-$\sigma$ detection limit for the methanol multibeam survey of the LMC (assuming a distance of 50~kpc).  The dashed sloped line shows the lower-limit on 6.7- to 12.2-GHz flux density ratio for these observations computed using the detection limit of the 12.2-GHz observations.}
  \label{fig:gal_lmc}
\end{figure}

There are a number of publications which give a detailed comparison of the properties of large samples of Galactic 6.7- and 12.2-GHz methanol masers, in particular \citet{Caswell+95} and \citet{Breen+09}.  Figure~\ref{fig:gal_lmc} shows a log-log plot of the ratio of the 6.7- to 12.2-GHz peak versus the 6.7-GHz isotropic luminosity (determined from the peak flux density and measured in units of Jy kpc$^2$) for a sample of 63 Galactic 12.2-GHz methanol masers \citep{Breen+09}.  Figure~\ref{fig:gal_lmc} includes the four LMC points as filled squares.  Our upper limits for the 12.2-GHz emission towards the 6.7-GHz methanol masers other than N105a, allows us to determine a lower limit to the 6.7- to 12.2-GHz peak flux ratio for these sources.  These lower limits are indicated by the presence of an upward pointing arrow.  Fig.~\ref{fig:gal_lmc} shows that the ratios (or limits) for 6.7- to 12.2-GHz emission in the LMC lie within the range observed in Galactic sources.  The dashed vertical line indicates the 5-$\sigma$ detection limit for the methanol multibeam survey of the LMC \citep{Green+08}.  The dashed sloped line indicates how the lower-limit on the 6.7- to 12.2-GHz peak flux density ratio changes as a function of 6.7-GHz peak luminosity for sources at the distance of the LMC for observations with a 5-$\sigma$ detection limit of 0.135~Jy (the limit for our 12.2-GHz observations).  Our observations, when combined with those of \citeauthor{Green+08} show that there are no methanol masers in the LMC with characteristics which place them in the lower-right region of Fig.~\ref{fig:gal_lmc} (as delineated by the dashed lines).  However, only 20 per cent of Galactic masers with both 6.7- and 12.2-GHz fall within these limits (which denote the region occupied by luminous 6.7-GHz methanol masers with a relatively strong 12.2-GHz counterpart). Two of the lower-limits established for the 6.7- to 12.2-GHz peak flux density ratio are well below the median value observed in Galactic sources for 6.7-GHz masers of comparable luminosity.  However, observations approximately two orders of magnitude more sensitive would be required to establish whether these sources are under-luminous at 12.2-GHz compared to similar Galactic sources.  For the most luminous LMC 6.7-GHz methanol maser {\em IRAS}\,05011-6815, the established lower limit shows that any 12.2-GHz methanol maser in this source is less luminous than for the majority of similar Galactic 6.7-GHz methanol masers, but well within the observed range.  In summary, with our existing data there is no evidence for any significant difference between the ratios of methanol maser lines seen in the LMC compared to the Milky Way.  Fig.~\ref{fig:gal_lmc} also nicely illustrates that the majority of the known 6.7-GHz methanol masers in the LMC are not particularly luminous in comparison with the luminosities of the Galactic sample shown here.

We can also compare the rate of association and relative strength of the 12.2-GHz methanol and ground-state, main-line OH masers with those observed in Galactic samples.  \citet{Caswell98} provides the most comprehensive, sensitive sample of Galactic ground-state OH masers in star forming regions, 80 per cent of which have an associated 6.7-GHz methanol maser.  There are a total of four ground-state OH masers known in the LMC \citep{Brooks+97},  and on the basis of Galactic statistics we would expect 3--4 of these to have associated 6.7-GHz methanol masers, but only two do ({\em IRAS}\,05011 \& N160a).  \citet{Breen+09} found that 12.2-GHz methanol masers are found preferentially towards 6.7-GHz methanol masers with an associated ground-state OH masers.  Our single detection does not follow this trend.

In sources where 1665-MHz OH, 6.7-GHz methanol and 12.2-GHz methanol are all observed, both the OH and 12.2-GHz methanol are, typically, significantly weaker than the 6.7-GHz methanol masers \citep{Caswell+95}.  \citet{Caswell+93} studied the relative intensity of 12.2-GHz methanol and ground-state OH lines in a sample of 173 OH masers.  They found a large scatter in the observed ratio, with a 12.2-GHz:OH median ratio of 5:1 for the 53 sources where both transitions were detected, falling to 1:1 if 12.2-GHz non-detections were included in the sample.  This suggests that where both transitions are observed the 12.2-GHz methanol masers will usually be stronger.

The small sample size for all the different maser transitions observed in the LMC, along with the wide range of intensity ratios observed in Galactic samples does not allow us to draw any firm conclusions as to whether the properties of the LMC masers differ from those seen our Galaxy.  However, the absence of 12.2-GHz methanol masers towards sources with both 6.7-GHz methanol and ground-state OH masers, and the rate of detected 6.7-GHz methanol towards OH masers in the LMC are both consistent with the established trend for the Magellanic Clouds to have an underabundance of methanol masers compared to the Galaxy.

Comparison of the relative rates of association between methanol and water masers is more difficult because less work has been done on this in the Milky Way.  At present there are no published large-scale unbiased water maser surveys to compare to the methanol maser surveys.  However, we note that this will change in the relatively near future when the results of the H$_2$O galactic Plane Survey (HOPS) are published \citep{Walsh+08}.  However, until the results of HOPS are available, the only information we have comes from targeted searches of one transition towards the other \citep[e.g.][]{Szymczak+05,Xu+08}.  \citet{Szymczak+05} found a detection rate of approximately 50 per cent for water masers towards known 6.7-GHz methanol masers.  Conversely \citet{Xu+08} detected 6.7-GHz methanol masers towards approximately 30 per cent of a sample of water masers.  Our observations show that of the four 6.7-GHz methanol masers in the LMC, two have an associated water maser ({\em IRAS}\,05011-6815 and N160a/MC76), a detection rate of 50 per cent.  Including the six new water masers we have detected, there are a total of thirteen water masers associated with star formation in the LMC.  Hence the detection rate of 6.7-GHz methanol masers towards water masers is 15 per cent for the LMC, significantly lower than in the Galaxy and again consistent with the observed under-abundance of methanol masers in the LMC.

\subsection{Masers as star formation tracers}

We have searched the {\em Spitzer} SAGE Infrared Array Camera (IRAC) point source catalogue and archive for sources within 5 arcseconds of the maser positions.  Four of the maser sources were associated (separation less than 5 arcsec) with sources in the SAGE IRAC catalogue. These are listed in Table~\ref{tab:ir}. Somewhat surprisingly none of these sources were identified as YSO candidates by \citet{Whitney+08} in their catalogue of candidate young stellar objects.  We have examined the Multiband Imaging Photometer (MIPS) 24-$\mu$m images from the SAGE program and find that all the maser sources are associated either with a point source (listed in Table~\ref{tab:ir}), or are projected against more extended mid-infrared emission.
  
\begin{table*}
\caption{The association of known interstellar masers in the LMC with infrared sources from SAGE.  {\bf NOTE:} In the N113/MC24 field 051317.69-692225.0 is more than 2 arcseconds away from two of the masers, and we have given these separations merely as useful information.}
  \begin{tabular}{llllrclrc} \hline
      \multicolumn{1}{c}{\bf Source} & \multicolumn{1}{c}{\bf Green} & \multicolumn{4}{c}{\bf SAGE Associations} &  \multicolumn{3}{c}{\bf Gruendl \& Chu} \\
     \multicolumn{1}{c}{\bf name}     & \multicolumn{1}{c}{\bf et al.} & \multicolumn{1}{c}{\bf IRAC} & \multicolumn{1}{c}{\bf MIPS} & \multicolumn{1}{c}{\bf Dist.} & \multicolumn{1}{c}{\bf YSO} & \multicolumn{1}{c}{\bf Association} & \multicolumn{1}{c}{\bf Dist.} & \multicolumn{1}{c}{\bf YSO} \\
     & \multicolumn{1}{c}{\bf name}  & & & \multicolumn{1}{c}{\bf $^{\prime\prime}$} & \multicolumn{1}{c}{\bf Cand.} & & \multicolumn{1}{c}{\bf $^{\prime\prime}$} & \multicolumn{1}{c}{\bf Cand.} \\ \hline \hline
   N11/MC18                          & LMCm01 & j045647.25-662433.6 & & 2.1 & N & & & \\
  {\em IRAS}\,05011--6815 & LMCm02 & j050101.79-681028.8 & j050101.83-681028.5 & 0.7,0.3 & N & 050101.80-681028.5 & 0.3 & Y \\
  N105a/MC23 (OH)             & LMCm03 & j050952.04-685327.1 & & 1.5 & N & 050952.26-685327.3 & 1.9 & Y \\
                                                 &                   & j050952.04-685327.1 & & 2.9 & N & 050952.26-685327.3 & 1.7 & Y \\
  N105a/MC23 (meth)          & LMCm04 & j050958.51-685435.2 & j050958.77-685436.6 & 1.3, 2.5 & N & 050958.52-685435.5 & 1.6 & Y \\
  N113/MC24			  & LMCm05 & & & & & 051325.09-692245.1 & 0.4 & Y \\
                                                 &                  & & & & & 051317.69-692225.0 & 4.4 & Y \\
   					  & LMCm06 & & & & & 051317.69-692225.0 & 3.6 & Y \\
     					  &                   & & & & & 051317.69-692225.0 & 2.0 & Y \\
  N157a/MC74			  & LMCm07 & & & & & & & \\
  					  &                   & & & & & 053849.27-690444.4 & 2.3 & Y \\
					  &                   & & & & & 053852.67-690437.5 & 0.3 & Y \\
                                                 & LMCm08 & & & & & 053845.15-690507.9 & 0.9 & Y \\ 
  N159				  & LMCm09 & j053930.06-694720.5 &  j053929.45-694718.6 & 4.6, 1.2 & N & 053929.21-694719.0 & 0.1& Y \\
  N160a/MC76			  & LMCm10 & j053943.77-693833.4 &  & 0.5 & N & 053943.82-693834.0 & 0.2 & Y \\
 			                     & LMCm11 & & & & & 053939.02-693911.4 & 0.7 & Y \\ \hline
  \end{tabular} \label{tab:ir}
  \end{table*}

\citet{Gruendl+09} have independently searched for YSO candidates in the LMC.  Their analysis, likewise, used the SAGE data, but also included a large amount of published and archival data from other wavelengths.  We have compared the positions of the interstellar masers in the LMC with their catalogue and find that 11 of the 16 locations are within 2 arcseconds of an object they have classified as a ``definite'' YSO. The sources without an associated YSO candidate all lie in regions where there is a complex background, or a bright source nearby (Robert Gruendl, private communication).  It is striking that the methods used by \citet{Whitney+08} and \citet{Gruendl+09} to identify YSOs in the LMC (with the bulk of the observational data in common) produce such very different results.  Clearly \citet{Gruendl+09} have been much more successful in identifying the young, high-mass objects with which masers are associated.

\begin{figure*}
   \begin{center}
   \begin{minipage}[t]{0.45\textwidth}
       \psfig{file=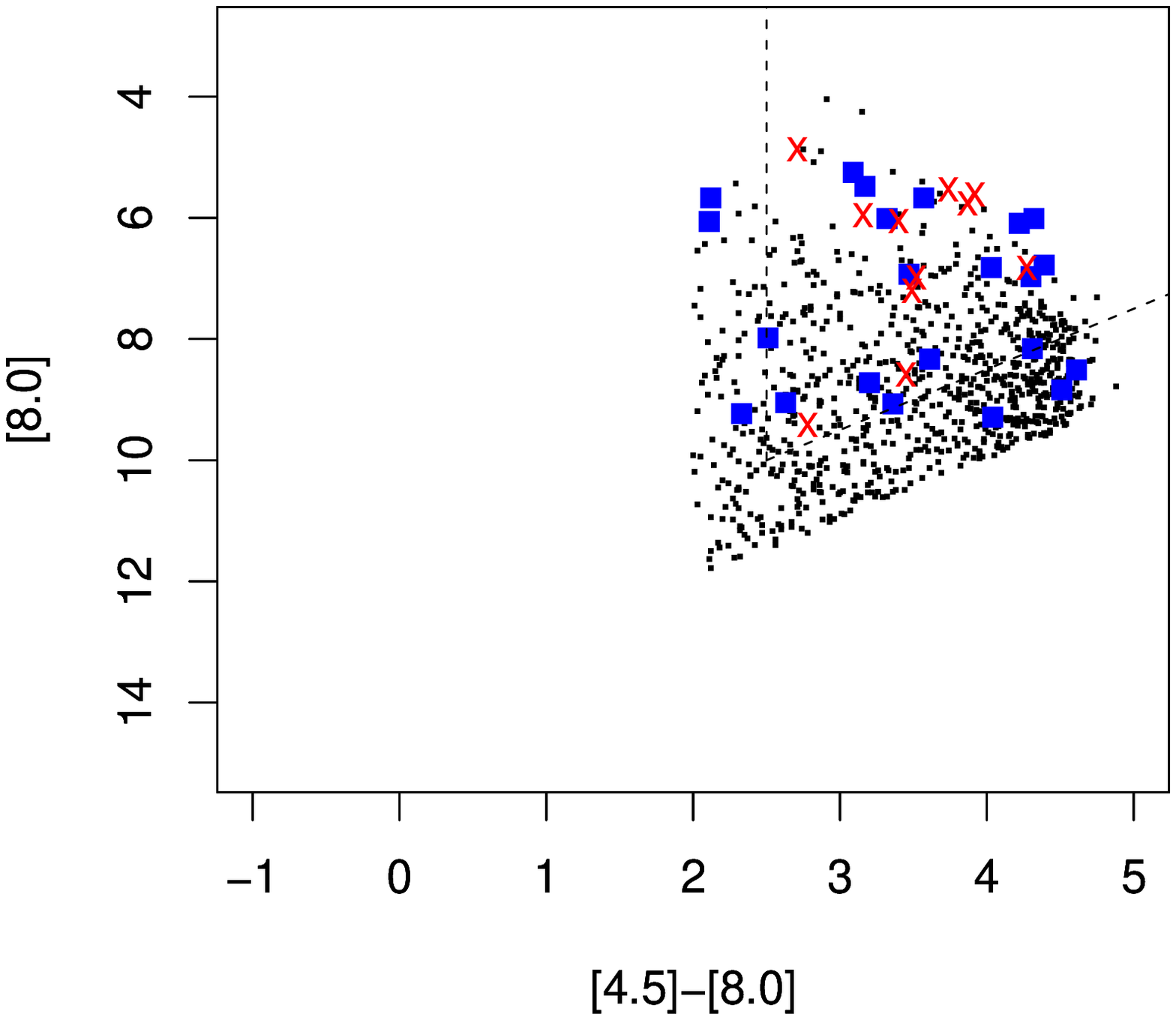,height=1.1\textwidth,width=1.1\textwidth}
   \end{minipage}
   \hfill
   \begin{minipage}[t]{0.45\textwidth}
       \psfig{file=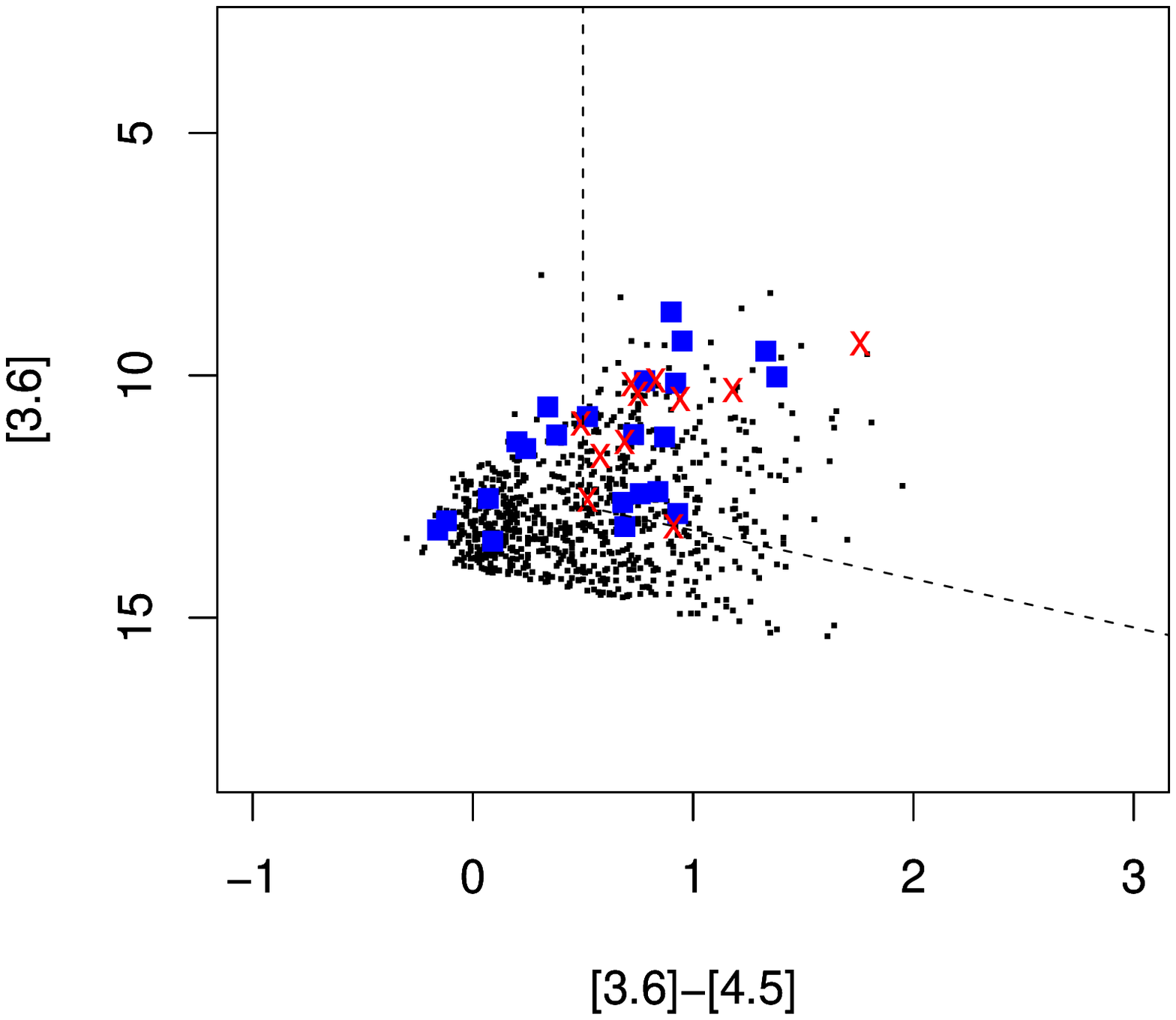,height=1.1\textwidth,width=1.1\textwidth}
   \end{minipage}
 \end{center}
  \caption{Colour-magnitude diagrams comparing the sample of definite YSOs in the LMC from \citet{Gruendl+09} (black points), with the sub-sample of sources with an associated interstellar maser (red cross) and those sources within the maser fields without an associated maser (blue squares).  The tendency for the maser-associated sources to be associated with more luminous and redder sources can clearly be seen.  \citet{Gruendl+09} used a [4.5]-[8.0] versus [8.0] colour-magnitude diagram as the initial basis for their investigations.  All the sources in their sample meet the criteria [4.5]-[8.0] $>$ 2 and [8.0] $<$ 14 - ([4.5]-[8.0]).  The dashed lines in each plot indicate the region where most of the maser sources lie.  For the left-hand plot this is [4.5]-[8.0] $\ge$ 2.5 and [8.0] $\le$ 12.5-([4.5]-[8.0]) and for the right-hand plot this is [3.6]-[4.5] $\ge$ 0.5 and [3.6] $\ge$ 12.2+([3.6]-[4.5]).}
  \label{fig:colmag}
\end{figure*}

We have compared the infrared properties of the maser-associated sources to those of the whole sample of definite YSOs identified by \citet{Gruendl+09} (their Table~9).  Our initial examination of infrared colour-colour and colour-magnitude diagrams suggested that the maser-associated YSOs tended to be brighter in all near and mid-infrared wavelength bands and generally have redder colours than the bulk of YSO candidates (Figure~\ref{fig:colmag}).  This is consistent with the maser-associated sources being more luminous and more deeply embedded than the bulk of YSO candidates, and may be useful for identifying regions of parameter space associated with the rare, and highly sought youngest high-mass stars.  The discrimination of the maser-associated sources is most clearly seen in the colour-magnitude diagrams, which explains why similar investigations undertaken with samples of Galactic masers typically produce less conclusive results.  For Galactic sources the broad range of distances to the sources adds additional scatter to the observed magnitudes, an effect which is difficult to correct for when the distances to most sources are poorly known.
For the [8.0] versus [4.5]-[8.0] colour-magnitude diagram the masers-associated sources all satisfy the criteria [4.5]-[8.0] $\ge$ 2.5 and [8.0] $\le$ 12.5-([4.5]-[8.0]).  A total of 352 of the 855 YSOs meet these criteria.  For the [3.6] versus [3.6]-[4.5] colour-magnitude diagram the maser-associated sources (with one exception) satisfy the criteria [3.6]-[4.5] $\ge$ 0.5 and [3.6] $\ge$ 12.2+([3.6]-[4.5]) and a total of 251 of the 855 YSOs in the LMC meet these criteria. The dashed lines in Fig.~\ref{fig:colmag} indicate these criteria for each colour-magnitude diagram.  Taking the union of the two sets results in a total of 195 YSOs satisfying both sets of criteria (22 percent of the sample as a whole).

\begin{figure*}
  \psfig{file=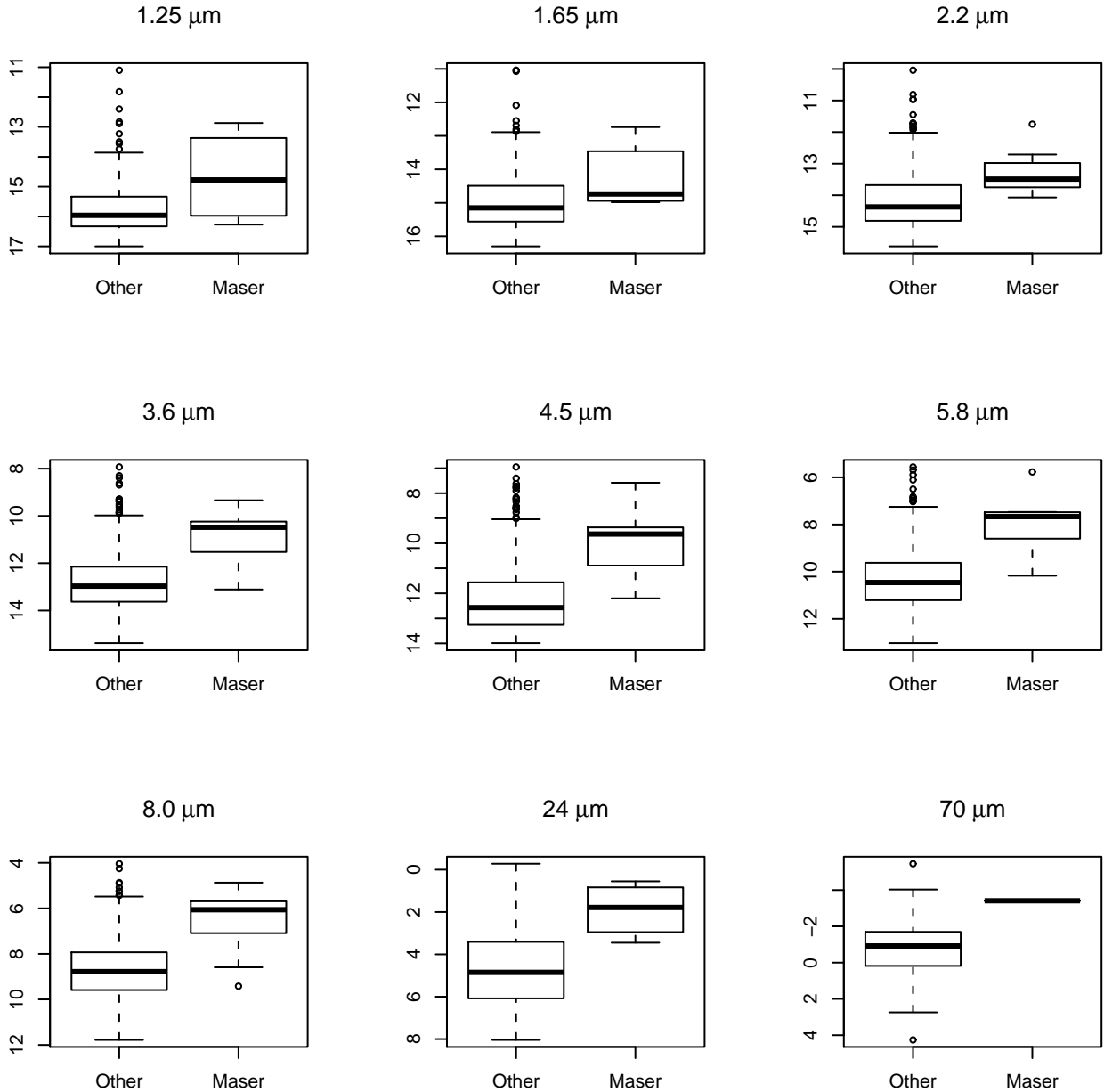}
  \caption{Boxplots comparing the distribution of infrared intensity (magnitude) for various wavelength bands for sources known to be associated with masers and for the rest of the sample (``Other'').  The y-axis for each box plot gives the intensity in magnitudes.  The solid line in the boxplot shows the median, the box shows the range between the 25$^{th}$ and 75$^{th}$ percentile, the vertical line from the top of the box shows the range from the 75$^{th}$ percentile to the maximum value and the vertical line from the bottom of the box shows the range from the 25$^{th}$ percentile to the minimum.  Extreme values (outliers) are represented by dots and are defined as points that are separated from the 25$^{th}$ or 75$^{th}$ percentiles by more than 1.5 times the interquartile range.  We observe a statistically significant difference in the mean of the maser and other distributions for all wavelength bands except 1.25~$\mu$m (J), 1.65~$\mu$m (H) and 70~$\mu$m.}
  \label{fig:boxplot}
\end{figure*}

Recent statistical investigations of millimetre dust clumps with and without associated masers has enabled the development of models which predict the presence of a maser on the basis of the dust-clump properties \citep{Breen+07,Breen+09}.  These models also allowed some inferences to be made about the critical physical properties which determine the presence of a maser.  Motivated by the apparent differences between those YSOs with, and those without masers in the colour-magnitude diagrams we decided to undertake a similar investigation using the infrared photometric data compiled by \citet{Gruendl+09} for the sample of YSOs in the LMC.  We constructed boxplots for each of the nine wavelength bands available (J, H, K from the Two Micron All Sky Survey, 3.6, 4.5, 5.8 and 8.0 $\mu$m from IRAC and 24 and 70 $\mu$m from MIPS) which show that in each of these wavelength bands the YSOs associated with masers tend to be more luminous than the sample as a whole (Figure~\ref{fig:boxplot}).  The photometric data in \citet{Gruendl+09} is only complete for the 4.5- and 8.0-$\mu$m bands, as these were used in the initial selection of YSO candidates.  While most sources have measurements for all of the IRAC wavelength bands, this is not the case for the 2MASS and MIPS data, especially for the maser-associated YSOs.  In order to objectively test whether the maser-associated YSOs are more luminous than the sample as a whole we performed a $t$-test for each wavelength band to determine if the two groups have a different mean.  The $t$-test returned a statistically significant result ($>$ 95 per cent probability that the means of the two samples are different) for each of the K-, 3.6-, 4.5-, 5.8-, 8.0- and 24-$\mu$m bands which gave $p$-values of 0.03, 0.0003, 0.0004, 0.0001, 0.0004 and 0.02 respectively.  For the other wavelength bands the number of maser-associated YSO with a measurement was in all but one case, less than half the sample (of 11 sources), which prevented meaningful statistical comparisons from being undertaken.  


We also investigated the infrared colours formed from the IRAC data.  From the four wavelength bands six independent colours can be formed.  The three colours which include data from the 3.6-$\mu$m band all have a statistically different mean (as measured by a $t$-test) for the maser-associated sources compared with the rest of the sample.  The $p$-values for the difference in the mean for the [3.6]-[8.0], [3.6]-[5.8] and [3.6]-[4.5] colours are 0.03, 0.03 and 0.01 respectively.  In each case the maser-associated sources have a larger mean colour (they are ``redder''), than the rest of the sample .

It would be desirable to fit a Binomial generalised linear model (GLM) to the infrared data to predict the properties of sources likely to be associated with a maser.  However, in order to undertake such modeling the only sources which can be used are those where the presence or absence of a maser is well established, and which have data for all the parameters.  For this reason, it was only practical to consider using the IRAC data for the GLM fitting as the data for the other wavelength bands are much less complete.  Of the 855 definite YSO sources identified by \citet{Gruendl+09} 814 have observations for all four IRAC wavelength bands.  Unfortunately, the small number of detected methanol masers and the targeted/incomplete nature of the water maser searches meant that for neither transition was GLM modeling practical due to the small sample sizes.

Fig.~\ref{fig:colmag} uses blue squares to represent those YSO within a maser field which do not have an associated maser.  It is clear that the majority of these sources have similar infrared properties to the maser associated sources.  This is in contrast to our findings when we compared the maser-associated YSOs to the rest of the sample.  When we perform $t$-tests comparing those YSOs which fall within a maser field with the rest of the sample we find a statistically significant difference between the means of the two samples for all the infrared bands with the exception of the MIPS 70~$\mu$m band (where there is insufficient data in the YSO near a maser sample).  In addition to this, the $p$-values from the $t$-tests are in all cases lower than when comparing the masers associated YSOs with the rest of the maser sample.  This is likely to be because the presence of a strong water maser signposts a young high-mass star formation region and since they form in clusters there are likely to be other similar sources nearby.  This is in contrast with most of the YSO sources, which differ significantly in their infrared properties from those associated with masers.  They are likely to either be from more-evolved regions or associated with low- or intermediate-mass star formation.  

In summary, while we have established that there is a statistically significant difference between the infrared properties of maser-associated YSOs compared to the whole sample, the information we have on the presence or absence of masers is insufficient to enable us to undertake predictive statistical modeling at the present time.

\subsection{SED modeling}
Colour-colour and colour-magnitude diagrams are useful for identifying regions of parameter space which favour certain types of objects.  They are always subject to a degree of confusion however, due to overlaps in the colour ranges of different types of objects \citep[as nicely demonstrated by ][]{Gruendl+09}.  In addition, they are often incomplete (sources need measurements in each of three or four wavelength bands to be included), or they do not utilize all the available information (for sources where there are measurements in other wavebands).  Fitting the spectral energy distribution (SED) of a source has the advantage that it can utilize all the available information for a particular source.  We have used the online fitting SED tool of \citet{Robitaille+07} to investigate the SED's of those YSOs associated with masers and also the properties of the other LMC YSOs.  The SED's are fitted to a grid of two-dimensional radiative transfer models of YSOs consisting of a central star surrounded by an accreting disk and envelope \citep{Robitaille+06}.  The limitations inherent in the assumptions, implementation and results of the model have been extensively discussed in the literature \citep[see for example][]{Robitaille08,Grave+09} and we will not repeat them here.  

\begin{figure*}
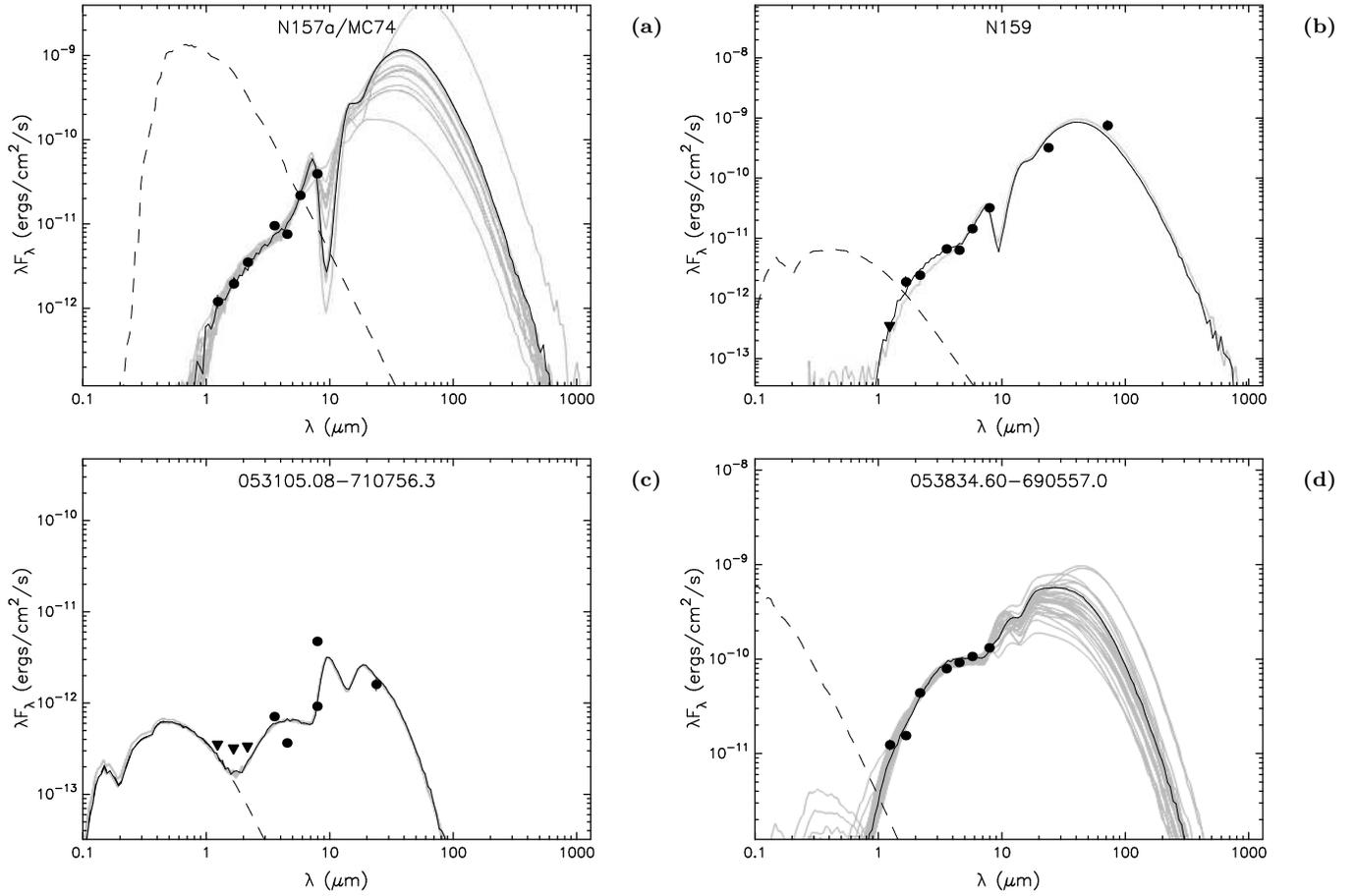

   \begin{center}
      \begin{minipage}[t]{0.45\textwidth}
         \psfig{file=fig9a.eps,width=1.0\textwidth}
     \end{minipage}
   \hfill
    \begin{minipage}[t]{0.01\textwidth}
       {\raisebox{55mm}{\bf{(a)}}}
     \end{minipage}
     \hfill
    \begin{minipage}[t]{0.45\textwidth}
       \psfig{file=fig9b.eps,width=1.0\textwidth}
   \end{minipage}
   \hfill
     \begin{minipage}[t]{0.01\textwidth}
     {\raisebox{55mm}{\bf{(b)}}}
   \end{minipage}
 \end{center}
 \begin{center}
   \begin{minipage}[t]{0.45\textwidth}
       \psfig{file=fig9c.eps,width=1.0\textwidth}
   \end{minipage}
   \hfill
     \begin{minipage}[t]{0.01\textwidth}
       {\raisebox{55mm}{\bf{(c)}}}
     \end{minipage}
     \hfill
   \begin{minipage}[t]{0.45\textwidth}
       \psfig{file=fig9d.eps,width=1.0\textwidth}
   \end{minipage}
   \hfill
     \begin{minipage}[t]{0.01\textwidth}
       {\raisebox{55mm}{\bf{(d)}}}
     \end{minipage}
 \end{center}
  \caption{The SED and fitted models for two maser associated sources (a \& b), a source with colours different from maser-associated YSOs (c) and a YSO within a maser field (d).  The circles represent measurements taken from Table~9 of \citet{Gruendl+09}, the triangles are upper limits for 2MASS 6x and SAGE observations of the LMC.  The best fit model is the solid black line and the grey lines show other acceptable models (chi-squared values within 3 of the best fit model).  The dashed line shows the spectrum of the stellar photosphere as it would appear in the absence of circumstellar dust, but including interstellar extinction.  The fitted mass of the central object and bolometric luminosity for each of these sources are (a) M=35.8\msol\/ ; L = $1.26 \times 10^5$\lsol\/ (b) M=28.9\msol\/ ; L = $1.16 \times 10^5$\lsol\/ (c) M=8.4\msol\/ ; L = $3.45 \times 10^3$\lsol\/ (d) M=18.4\msol\/ ; L = $9.34 \times 10^4$\lsol\/.}
  \label{fig:sedfit}
\end{figure*}

In order to fit each of the 855 definite YSO sources of \citet{Gruendl+09}, we used an automated script to populate the online fitting form with the infrared data and download the best-fit model parameters for each source.  We have used the photometric measurements of \citet{Gruendl+09} for each of the bands where they are available.  When there was no reported measurement for a particular waveband we have assumed upper limits of 17.6, 16.9 and 16.1 magnitudes for the J-, H- and Ks-band deep 2MASS observations towards the LMC \citep{Cutri+02} and upper limits of 19.3 and 16.1 magnitudes for the IRAC 3.6- and 5.8-$\mu$m bands (the observations are complete in the other IRAC bands due to the selection criteria) \citep{Sewilo08}.  Where no measurement was available for the MIPS 24- and 70-$\mu$m bands we chose not to use the upper limit values in the fit, as in many cases the sources are in regions of extended or saturated emission at these wavelengths, rather than regions with no emission.  Following \citet{Robitaille+07} we have assumed aperture sizes of 3 arcseconds for the 2MASS observations, 5 arcseconds for IRAC observations, 10 arcseconds for MIPS 24~$\mu$m and 20 arcseconds for MIPS 70~$\mu$m.  We allowed the distance to vary between 48 and 52~kpc and the visual extinction (along the line of sight) to vary between 0 and 2 magnitudes (the mean value for the interstellar extinction towards the LMC measured by \citet{Imara+07} is 0.3 magnitudes).  Fig.~\ref{fig:sedfit} shows some examples of the resulting SED fits for maser and non-maser associated YSOs.

\begin{figure}
  \psfig{file=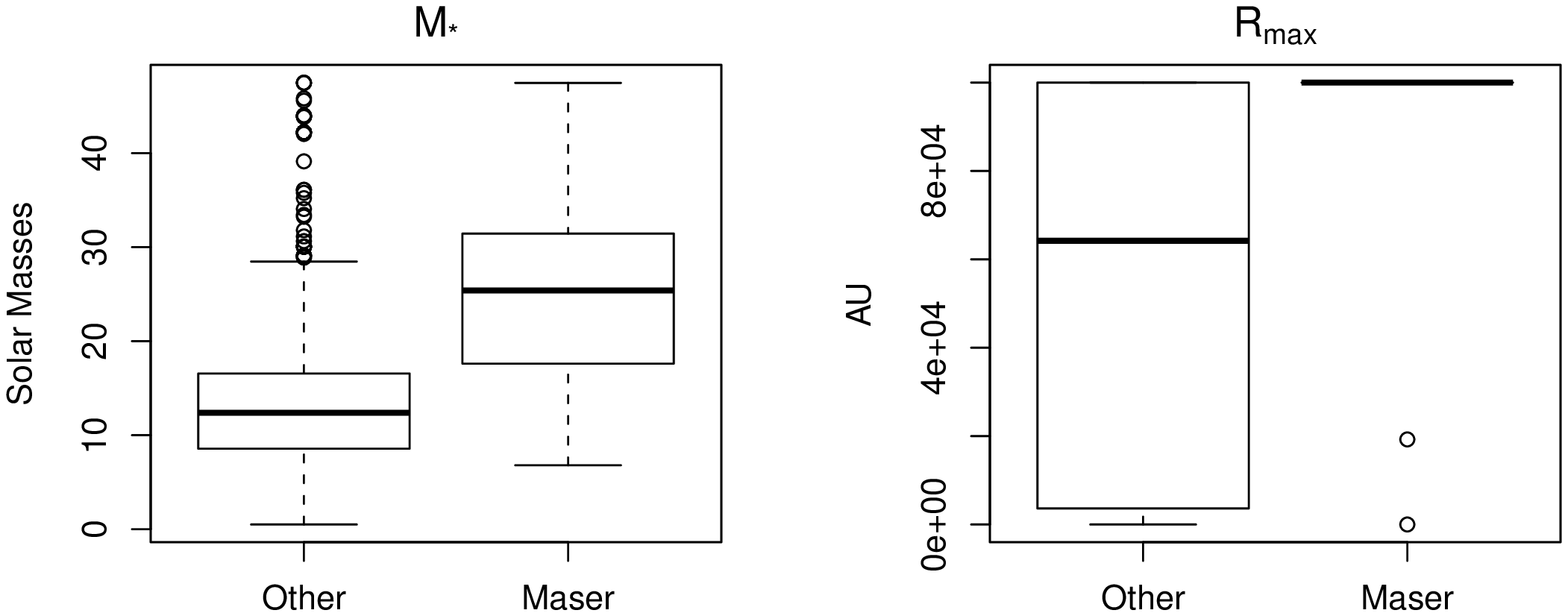,width=0.225\textwidth,angle=270}
  \psfig{file=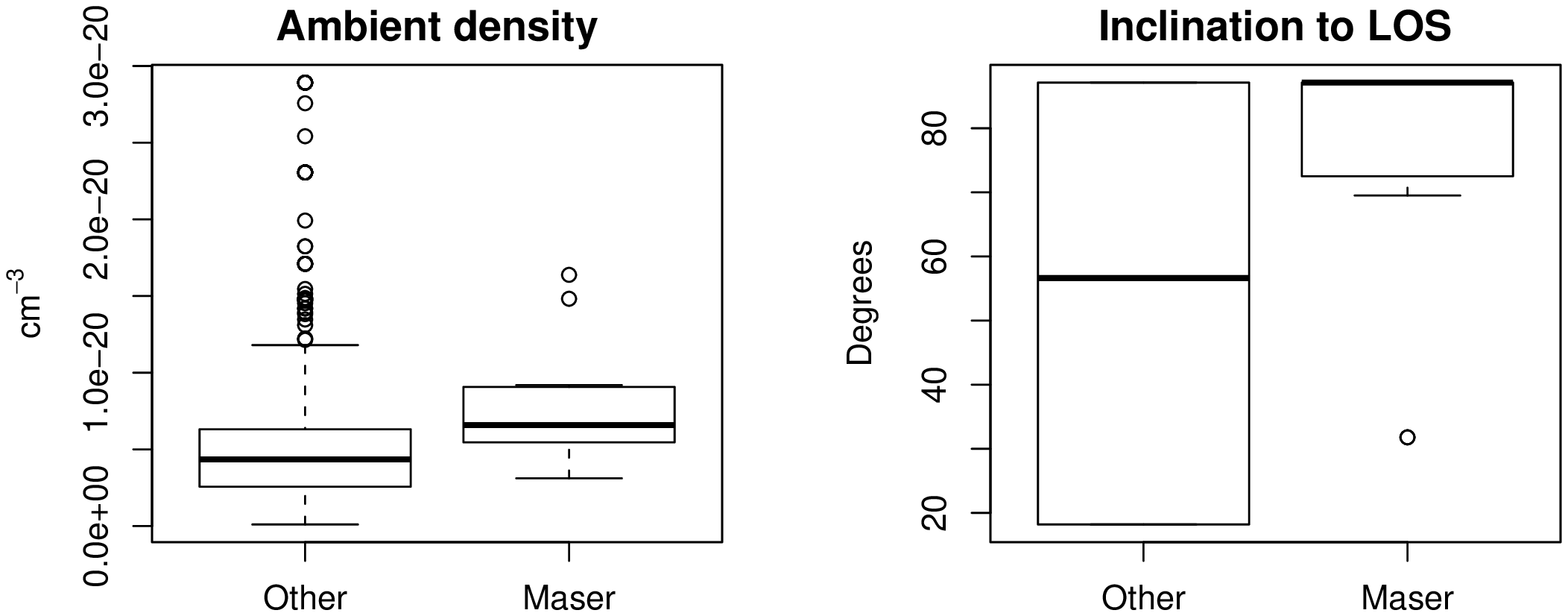,width=0.225\textwidth,angle=270}
  \psfig{file=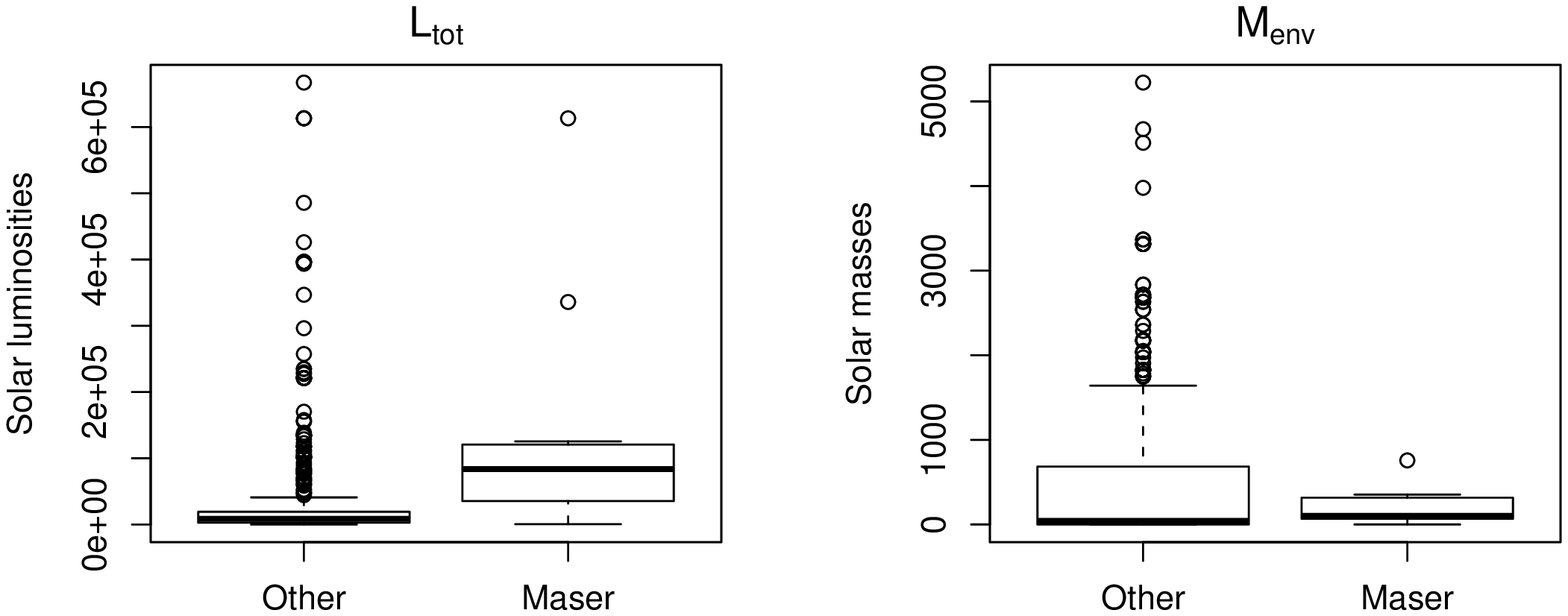,width=0.225\textwidth,angle=270}
  \caption{Boxplots comparing the distribution of some of the fitted parameters for the SED models for sources know to be associated with masers and for the rest of the sample (``Other''). The six parameters shown here are the five for which a Mann-Whitney test shows a statistically significant difference in the median of the two distributions, plus the distribution of the envelope mass.  See the caption for Fig.~\ref{fig:boxplot} for a description of how the boxplots can be interpreted.}
  \label{fig:sed_box}
\end{figure}

There are between seven and nine infrared intensities or upper limits available for each source and in most cases the chi-squared for the resulting SED fits are reasonable.  However, because of the large number of sources modeled we have made no attempt to remove sources where this was not the case. Figure~\ref{fig:sed_box} shows the distribution of some of the SED model fit parameters for the YSOs associated with masers compared to the other YSOs in the \citet{Gruendl+09} sample.  For most of the parameters there is a large range present in both categories of YSO and in many cases the distributions are highly skewed (e.g. the envelope mass shown in Fig.~\ref{fig:sed_box}).  For skewed distributions it is more appropriate to use the Mann-Whitney test to determine if the two samples have a different median, rather than the $t$-test (which compares the means).  The Mann-Whitney test returns statistically significant results ($p$-value $<$ 0.05) for five of the model parameters, the mass of the central star (0.001), the outer radius of the envelope (0.05), the ambient density (0.01), the inclination to the line of sight (0.01) and the total luminosity (0.0008).  The inferences which can be drawn from these results is clear.  The masers are associated with high-mass, luminous YSOs where the SED is dominated by a large envelope and are located in regions where the ambient density is higher than in the vicinity of most YSOs.  

The outer radius of the envelope and the inclination to the line of sight the results appear to be influenced by details of the modeling process.  The maximum outer radius of the envelope used by the model is $10^5$ AU, and that is the fitted value for 9 of the 11 maser-associated YSOs (compared to 360 of the other 844 sources).  Essentially this result is consistent with the masers being associated with higher mass objects, which would be expected to arise from the largest clumps/envelopes of gas and dust.  Similarly for the inclination angle the maximum value available in the models is 87.13$^\circ$, and that is the fitted value for 7 of the 11 maser-associated YSOs (compared to 252 of the other 844 sources).  Naively, it seems unlikely that geometry of the star forming region should play a significant role in whether we see a luminous water maser associated with a particular source and such a result is likely to be an artifact of the modeling process.  However, water masers are typically associated with outflows and an outflow-axis near perpendicular to the line of sight will produce the maximum line-of-sight velocity coherent path length and hence the most luminous water masers.  \citet{Robitaille+06} suggest that the inclination angle is one of the most important parameters in determining the SED for young objects, as it controls the contribution from the inner, hotter regions of the envelope due to orientation of the bipolar cavity.  The preference for the maser-associated YSOs to best fit models with high inclination angles then means that there is little or no contribution from the inner regions in these SEDs.  An alternative interpretation of these results would be that additional obscuring material is present for the highest mass YSOs in comparison to lower-mass objects.

We have compared the SED model parameters of the maser-associated YSOs with other YSOs in the maser fields.  We applied the Mann-Whitney test for all of the model parameters to compare these two samples and the only statistically significant result ($p$-value 0.03) was for the inclination to the line of sight.  The fact that the inclination angle again shows a statistically significant difference warrants further future investigation to determine whether it is an artifact of the modeling process, or if it has a physical basis.  Since the SED model fits are based on the same infrared data for which we found no significant difference between the maser-associated YSOs and those in the maser fields, it it not surprising that we see a similar lack of difference in the SED models.  This means that, as for the infrared properties it is not possible to undertake GLM modeling to try and statistically predict which of the other YSOs in the LMC may have an associated luminous water maser.  The best we can do is to look for other YSOs for which the fitted models are very similar to those associated with masers.  Of the list of 844 YSOs in the LMC which are not known to have an associated water maser there are a total of 17 for which each of the mass of the central star, total luminosity, ambient density, outer radius of the envelope and inclination to the line of sight are all greater than or equal to the median of the maser-associated YSO sample.  These YSOs are 044927.29-691206.3, 045358.57-691106.6, 045426.06-691102.3, 051004.00-685438.3, 051328.26-692241.8, 051922.98-693940.5, 052208.52-675922.1, 052257.55-680414.1, 053205.89-662530.9, 053603.03-673234.2, 53831.62-690214.6, 053912.67-692941.4, 053937.04-694536.7, 053937.53-694609.8, 053945.94-693839.2, 054309.58-675006.6 \&  054630.37-693458.6.  Three of these (051004.00-685438.3, 051328.26-692241.8, 053945.94-693839.2) are within one of the maser fields and so are known not to have an associated maser, however, the remaining 14 should be primary targets in any future searches for water masers in the LMC.

\section{Conclusion}

A sensitive search for 12.2-GHz methanol masers towards a sample of eight fields containing all currently known methanol, OH and water masers in the Large Magellanic Cloud has resulted in the detection of one source.  This is the first detection of a 12.2-GHz methanol maser outside our Galaxy.  Our sample size is too small and our observations have insufficient sensitivity to determine if there is any difference in the properties of the different maser transitions in the LMC compared to the Galaxy; however they are consistent with the previously observed underabundance of methanol masers in the LMC.

We have also observed the water masers associated with known high-mass star formation regions in the LMC and discovered six new sites.  Five of these are in regions where emission had previously been detected, while the sixth is a weak new water maser detected towards the strongest 6.7-GHz methanol maser in the LMC {\em IRAS}\,05011-6815. We find most of the star-formation masers in the LMC for which accurate positions are available are associated with sources identified as YSO on the basis of their infrared properties.  The YSOs associated with masers tend to be more luminous and redder than the sample as a whole.  We suggest that this is likely due to the masers being associated with young high-mass star formation regions.  SED modeling of the YSOs shows that the masers are preferentially associated with sources with a high central mass, total luminosity and ambient density.  This is consistent with expectations from Galactic maser studies.

\section*{Acknowledgments}

We would like to thank the referee Michele Pestalozzi for useful suggestions which have improved the paper.  We thank John Whiteoak and Jasmina Lazendic-Galloway for useful discussions on the nature of the water maser emission in N113/MC24.  We would like to thank Robert Gruendl for making available information on YSO candidates in the LMC prior to publication. We would like to thank Karl Menten for useful discussions during the writing of this paper.  SPE would like to thank the Alexander-von-Humboldt-Stiftung for an Experienced Researcher Fellowship which has helped support this research. This research was supported by the EU Framework 6 Marie Curie Early Stage Training programme under contract number MEST-CT-2005-19669 ``ESTRELA''.  The Parkes telescope and Australia Telescope Compact Array are part of the Australia Telescope which is funded by the Commonwealth of Australia for operation as a National Facility managed by CSIRO. This research has made use of NASA's Astrophysics Data System Abstract Service.  This research has made use of the NASA/IPAC Infrared Science Archive, which is operated by the Jet Propulsion Laboratory, under contract with the National Aeronautics and Space Administration.  This work is based in part on observations made with the Spitzer Space Telescope, which is operated by the Jet Propulsion Laboratory, under contract with the National Aeronautics and Space Administration.

\end{document}